\documentclass[aps,pre,preprint,superscriptaddress,showkeys]{revtex4-1}

\usepackage{graphicx}
\usepackage[absolute]{textpos}

\begin{document}
\title{How images determine our visual search strategy}

\author{Tatiana A. Amor}
\affiliation{Computational Physics IfB, ETH Zurich, Stefano-Franscini-Platz 3, CH-8093, Zurich, Switzerland}
\affiliation{Departamento de F\'{i}sica, Universidade Federal do Cear\'{a}, 60451-970, Fortaleza, Cear\'{a}, Brazil}
\author{Mirko Lukovi\'{c}} 
\affiliation{Computational Physics IfB, ETH Zurich, Stefano-Franscini-Platz 3, CH-8093, Zurich, Switzerland}
\author{Hans J. Herrmann}
\affiliation{Computational Physics IfB, ETH Zurich, Stefano-Franscini-Platz 3, CH-8093, Zurich, Switzerland}
\affiliation{Departamento de F\'{i}sica, Universidade Federal do Cear\'{a}, 60451-970, Fortaleza, Cear\'{a}, Brazil}
\author{Jos\'{e} S. Andrade, Jr.}
\email[Correspondence: ]{soares@fisica.ufc.br}
\affiliation{Departamento de F\'{i}sica, Universidade Federal do Cear\'{a}, 60451-970, Fortaleza, Cear\'{a}, Brazil}

\begin{abstract}
When searching for a target within an image our brain can adopt different
strategies, but which one does it choose? This question can be answered by tracking
the motion of the eye while it executes the task. Following many individuals
performing various search tasks we distinguish between two competing strategies.
Motivated by these findings, we introduce a model that captures the interplay of the
search strategies and allows us to create artificial eye-tracking trajectories,
which could be compared to the experimental ones. Identifying the model parameters
allows us to quantify the strategy employed in
terms of ensemble averages, characterizing each experimental cohort. In this way we
can discern with high sensitivity the relation between the visual landscape and the
average strategy, disclosing how small variations in the image induce
changes in the strategy.
\end{abstract}


\maketitle

\section*{Introduction}
What visual strategy do we employ when searching for a familiar face in a crowd and how would it change if we had to find a friend in a well organized choir? For instance,
do we explore each face sequentially or do we pick them randomly until reaching our target.
Along these lines, it is not at all obvious 
whether there exists a characteristic strategy that is related to the scene content. 
To shed light into this problem 
it is necessary to find a method that allows us to identify and quantify
particular features associated to the strategy adopted while searching for a hidden target.

Different models have been developed with the goal of understanding what guides eye movement during visual search (see Ref.~\cite{eckstein2011visual} for a review). One family of these models is based on the construction of \textit{saliency maps}, which define regions of interest derived from properties of scene objects (such as luminescence, color, orientation)~\cite{itti1998model,itti2000saliency,over2007coarse,foulsham2008can,ehinger2009modelling,nakayama2011situating}. By definition these salient regions stand out from other parts of the scene and are therefore more susceptible to frequent eye fixations. 
In the context of visual search, these models prove to be more suitable for tasks involving a small number of equally relevant \textit{distractors}, which are essentially all items in the scene that are not targets of the search~\cite{li2002saliency,shiffrin1972visual}. On the other hand, in more complex visual search tasks, the salient regions might not necessarily be relevant.
In order to address this issue, other implementations of the saliency model consider the relative information of an object with respect to the global information of the scene~\cite{zhang2008sun,bruce2009saliency,gao2009discriminant}.

In the absence of salient elements, it is not possible to use these models as there are no  \textit{a priori} privileged regions within the scene.
As a consequence, another family of visual searching models have been proposed, namely, the \textit{saccadic targeting models}~\cite{rao2002eye,beutter2003saccadic,eckstein2006attentional,pomplun2006saccadic,zelinsky2008theory}. In these studies, the main hypothesis is that saccadic eye movements are directed to locations within the scene that contain elements similar to the target.
This similarity can be due to the image content as well as to a neurobiological filter. Within this framework, Najemnik \& Geisler propose a model where each point in space has a certain probability of being explored and the saccadic movement is directed to the most probable regions~\cite{najemnik2005optimal,najemnik2008eye}. These probabilities are then updated over time so that regions that were already explored are less likely to be revisited, introducing in a natural way the notion of persistence while searching. Moreover, other implementations of this model have taken into account the proximity between consecutive saccadic movements by adding a cost function that punishes longer saccades~\cite{araujo2001eye}.

Scanpaths produced by eye movements in general reveal different forms of persistence, at the level of saccades~\cite{amor2016persistence} and also within the fixations~\cite{engbert2011integrated,marlow2015temporal}. During most forms of random search, persistence of the jittering movements (i.e. saccades in visual search) plays a crucial role since it 
unveils
the strategy involved~\cite{codling2008random,tejedor2012optimizing}. In fact, while looking for a hidden target in a field of distractors, a variety of patterns have been documented~\cite{amor2016persistence,credidio2012statistical,le2016introducing,motter1998guidance}, ranging from systematic or completely persistent to random. However, while some of the models described above do make use of some form of persistence, they do not relate it to the overall strategy of the search. Beside some exceptions~\cite{brockmann2000ecology,boccignone2004modelling}, the saliency models do not focus on understanding the sequence in which the fixations are performed but instead determine the regions where they are more likely to occur. On the other hand, the saccadic targeting models, although dealing with the saccadic sequence, do not account for different strategies. 

Here we propose a visual search model (VSM) 
that 
quantifies the global persistent behavior and the overall strategy employed while looking for a hidden target. 
The parameters of the VSM define the saccadic orientation distribution, which has experimentally proven to provide information regarding the strategy~\cite{amor2016persistence} and the scene content~\cite{le2016introducing}. 
By studying this distribution obtained from experimental data, we were able to identify different search strategies that emerge from exploring the same scene and quantify them through the VSM parameters.
We analyzed the search strategies adopted in three different visual search scenes that differ in the features and arrangement of the scene items.
We found that the average strategy changes with these scenes suggesting that scene content and structure influence the way subjects execute their search.

In what follows we first describe the VSM and how to extract the model parameters from the experimental scanpaths in the \textit{``cloud number} (CN)\textit{''} experiment. In this visual search task the participants are requested to find a unique number ``5'' embedded in a field of numbers ranging from ``1'' to ``9'' serving as distractors. By means of an efficiency measure about the eye paths, we validate our model with respect to the experimental data. Then, we identify the location of the experimental trajectories for three different visual search tasks in the model parameter space and compare the average strategies applied for each task. Given that the average position within the parameter space differs for the three cohorts, the VSM can serve as a tool to predict the average strategy of similar experiments.
Finally, we discuss the implications of our results in the frame of visual search and the possibilities of exploiting the model to be used in other fields. 

\begin{figure*}[!htb]

\includegraphics[width=1\textwidth]{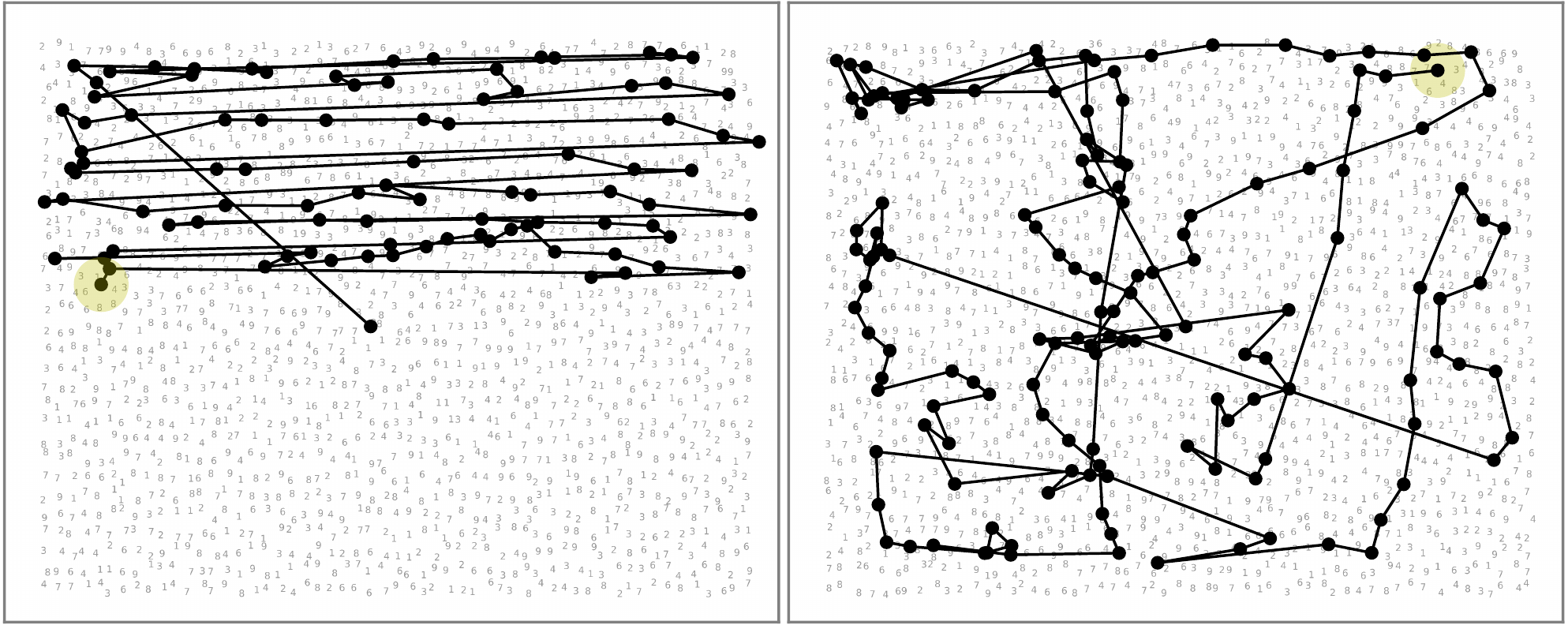}

\caption{\label{fig:fig_1} \textit{Different types of visual trajectories are found in the ``Cloud number (CN)'' experiment.} During a visual searching task, where participants have to find a unique number ``5'' embedded in a cloud of numbers between ``1'' and ``9'', participants perform very different trajectories. These trajectories range from a systematic search (\textbf{Left}) to a trajectory resembling more a random search with a strong persistence imprint (\textbf{Right}).}
\end{figure*}

\section*{Visual search model (VSM)}

While searching for a hidden target during a visual search episode,
participants perform different search strategies that vary from a very systematic to a seemingly random, as depicted in Fig.~\ref{fig:fig_1}. These strategies  are identified by studying the relative orientation between saccadic movements performed while searching. The saccadic relative angle distribution 
provides
information on the type of strategy employed~\cite{amor2016persistence} and may as well provide insight regarding the scene an observer is looking at~\cite{le2016introducing}. Therefore, the aim of the VSM is to emulate visual search trajectories via the study of the distribution of the inter-saccadic angles as main ingredient.  

\begin{figure}[!htb]
\centering
\includegraphics[width=0.6\textwidth]{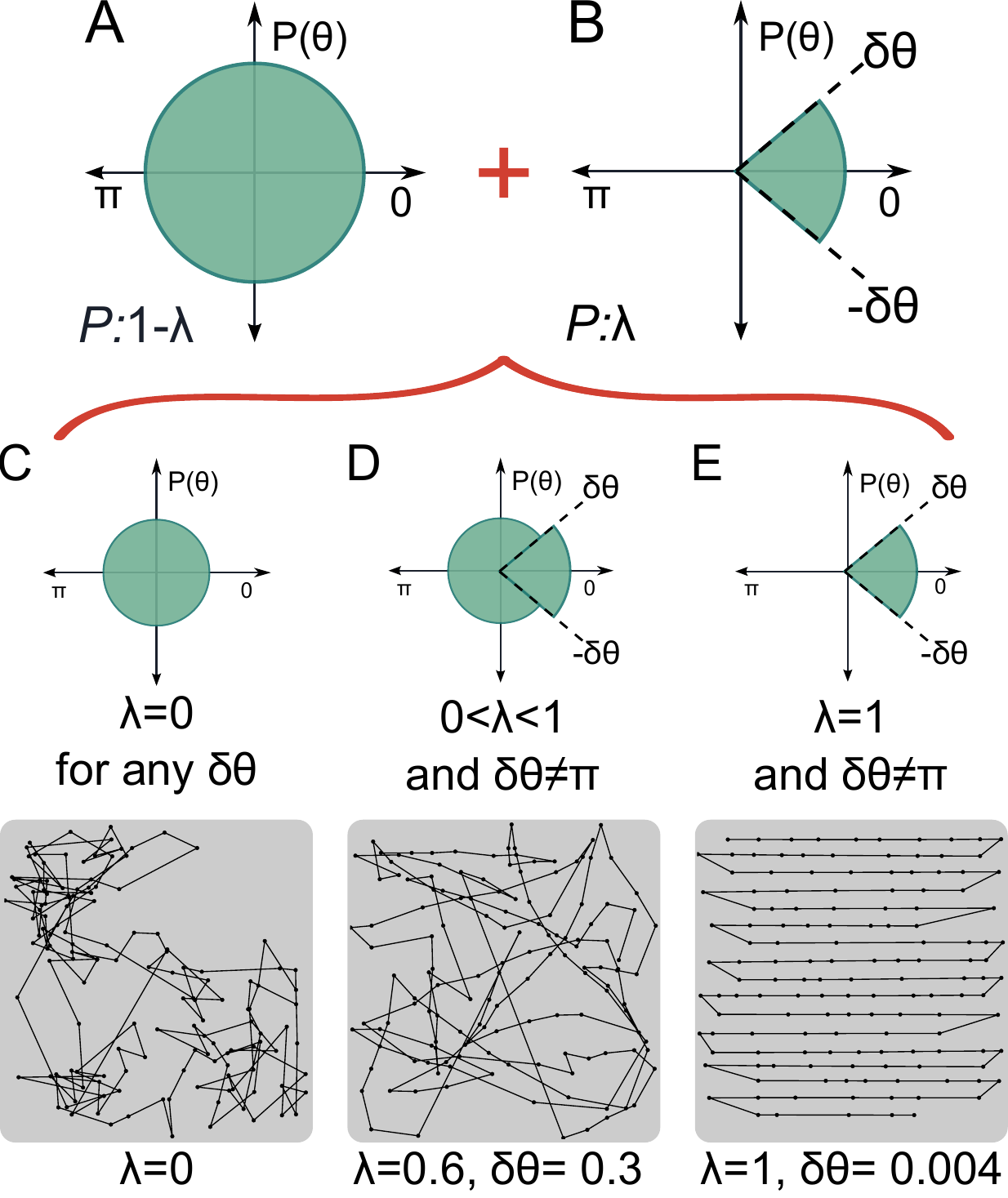}
\caption{\label{fig:fig_2} \textit{Schematic definition of the Visual Search Model (VSM).} Two parameters, $\lambda$ and $\delta\theta$ govern the way to select the relative angles between two consecutive saccadic movements. 
We consider two uniform distributions: one over all
possible angles (\textbf{A}) and one in the range $[-\delta\theta,\delta\theta]$ (\textbf{B}). With probability $1-\lambda$ angles are chosen from the first distribution and with probability $\lambda$ from the second one. Different combinations of these parameters result in a variety of distributions (\textbf{D}) responsible for different visual search strategies, ranging from a pure random walk (\textbf{C}) to a systematic search (\textbf{E}).}
\end{figure}

We sample the relative angle, $\theta$, between two consecutive saccadic movements from two distributions, as exemplified in Figs.~\ref{fig:fig_2}.A,B. The first distribution corresponds to a uniform distribution of all possible angles between $0$ and $2\pi$ (see Fig.~\ref{fig:fig_2}.A), and the second one is a uniform distribution with angles restricted to the interval $[-\delta\theta,\delta\theta]$ (see Fig.~\ref{fig:fig_2}.B).
Angles are chosen from the second distribution with probability $\lambda$, and from the first distribution with probability $1-\lambda$.
Therefore, only two parameters, $\lambda$ and $\delta\theta$, govern the statistical properties of the relative angle distribution.
The combination of these parameters gives rise to a wide range of different trajectories.
In the case $\lambda=0$, angles can only be sampled from the first distribution, corresponding to a purely random strategy, (see Fig.~\ref{fig:fig_2}.C). On the other hand, if $\lambda=1$, the angle is only sampled from the step distribution, so that only angles in the range $[-\delta\theta,\delta\theta]$ are allowed. If in the latter case we further assume that $\delta\theta$ is very small, so that each saccadic movement persists in the direction of the previous one, then the emergent trajectory will correspond to a systematic type of search (see Fig.~\ref{fig:fig_2}.E).

Between the two extreme cases, we find a repertoire of trajectories that result from the interplay between the parameters $\lambda$ and $\delta\theta$ (see Fig.~\ref{fig:fig_2}.D). The parameter $\lambda$ is related to the strategy employed during the search and  $\delta\theta$ to the persistent behavior of the search. 
For instance, an intermediate value of $\lambda$ combined with a small value of $\delta\theta$ results in a trajectory that can be identified as a random search with a strong persistence imprint.

Besides the relative angle distribution, another important ingredient in our model is the length of the saccadic movements, i.e the length of the jumps. We consider them to be sampled from a log-normal distribution, as reported in Ref.~\cite{credidio2012statistical}.

\vspace{0.2cm}
\noindent
\textbf{Estimation of $\lambda$ and $\delta\theta$ from the experimental data.}
From the VSM we propose the inter-saccadic relative angles, $\theta$, to be sampled from a combination of a uniform distribution over all angles and one in the interval $[-\delta\theta,\delta\theta]$ (see Fig.~\ref{fig:fig_2}). 
A conventional method to acquire the parameters from experimental data would be to study the histogram of $\theta$. However, 
one issue of this approach is 
that the parameter $\delta\theta$ is bin-size dependent and thus, is not a robust method for the parameter estimation. Moreover, the average number of $\theta$ values obtained in each trajectory is around 250, thus making it difficult to determine with precision the distribution from each of the scanpaths. 
One way around this issue is to
extract the model parameters directly from the experimental data. When the angles are ranked in increasing order from $-\pi$ to $\pi$, they present a characteristic shape of three linear segments, as shown in Fig.~\ref{fig:fig_3}.A with green dots. This shape is similar for all the trajectories (see Fig.S1).
In the extreme case that $\theta$ is sampled from the uniform distribution between $-\pi$ and $\pi$, the ranked angles form a straight line with a slope equal to $2\pi$.
On the other hand, if $\theta$ is sampled from the uniform distribution
between $-\delta\theta$ and $\delta\theta$, the ranked angles form a straight line where $\theta_{max}-\theta_{min}=2\delta\theta$. 
Considering the evident threefold behavior of the ranked angles obtained from the experiments and its origin as a combination of these three extreme cases, we fit the data using three consecutive linear segments (see Fig.~\ref{fig:fig_3}.A, solid orange lines). The middle region
depends on $\delta\theta$, and the first and third regions are 
related to the uniform distribution between $-\pi$ and $\pi$.  
The estimation of $\delta\theta$ is straightforwardly performed as the difference between the angle values defining the middle section
equals $2\delta\theta$.
The parameter $\lambda$ is the ratio between the number of angles originated by the oriented distribution, $N_{\lambda}$,
and the total number of angles, $N$. 
Within this framework, $N_{\lambda}$ corresponds to the difference between the total number of angles and $n_{step}$, defined as the angle index value for which the extension of the first linear segment is equal to $\pi$. 
This allows us to estimate the parameters of the VSM using the experimental data without the need of any intermediate manipulation.

\section*{Results}

\begin{figure*}[ht!]

\includegraphics[width=0.75\textwidth]{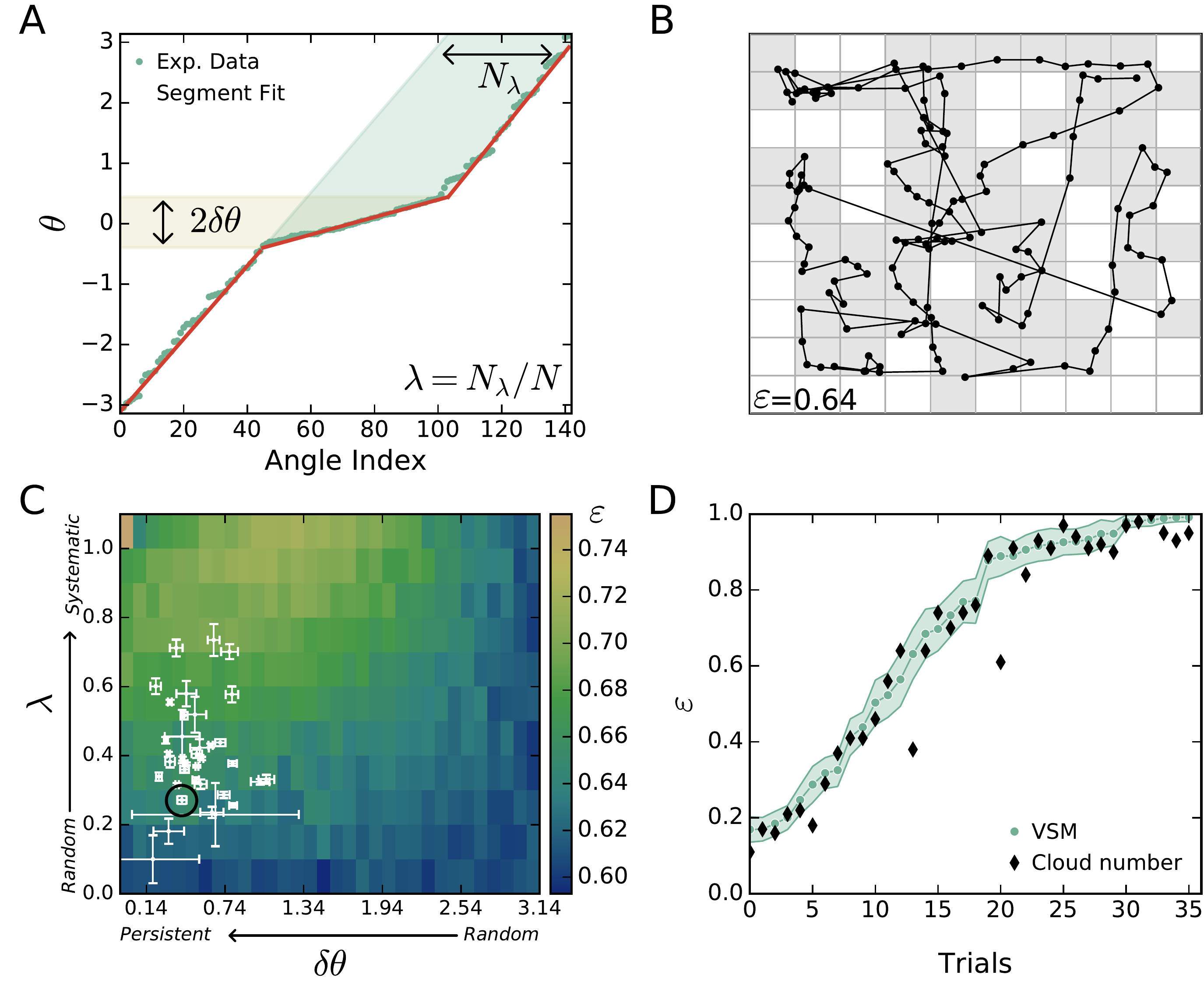}
\caption{\textit{Comparison of the VSM with the experimental eye paths.} \textbf{(A)} Estimation of $\lambda$ and $\delta\theta$ from the experimental trajectories. We rank $\theta$ (circles) for each experimental trajectory and perform a fit of three linear segments (solid lines). The limits of the middle region (yellow shaded region) define $\delta\theta$. The difference between the extended line of the first region and the end of the third region (green shaded region) defines the amount of angles, $N_{\lambda}$, originated by the distribution restricted by $\delta\theta$. The parameter $\lambda$ is defined as the ratio between $N_{\lambda}$ and the total number of angles $N$. \textbf{(B)} Definition of space filling efficiency, $\varepsilon$. $\varepsilon$ corresponds to the percentage of visited locations (gray cells) after parceling the image into $m\times m$ cells. In this case, $m=10$.  \textbf{(C)} Mean efficiency values obtained from the VSM parameters space for 100 realizations of the model. The experimental data from the CN cohort (white dots) are placed on their corresponding $\lambda$ and $\delta\theta$ values. \textbf{(D)} Comparison between $\varepsilon$ values obtained with the VSM and the experimental ones. For each trajectory (i.e black points) we run the model with the same number of jumps and the estimated $\lambda$ and $\delta\theta$ values. The shaded area corresponds to the standard deviation of 100 realizations of the model. The data presented in \textbf{(A)} and the data point highlighted in \textbf{(C)} correspond to the trajectory depicted in \textbf{(B)} with length $\sim140$. 
}

\label{fig:fig_3}
\end{figure*}

In order to compare the VSM with the experimental data
we study
the efficiency,  $\varepsilon$, defined in terms of
the space-filling attributes of the trajectories investigated (similar to the method applied in Ref.~\cite{engbert2006microsaccades})
The underlying hypothesis is that if the trajectory is able to fill the image space 
then the chances of finding the target increases, as more locations are examined.
The efficiency is measured in the following way: given a trajectory of a finite amount of steps, $N$,
we divide the figure being explored with a grid of $m\times m$ cells. Next, we determine the percentage of cells that were explored by the trajectory, as shown in Fig.~\ref{fig:fig_3}.B. When a fixation point falls into a cell, this cell is considered to be explored, since it is during fixational events that most visual information is gathered~\cite{hubel1995eye}. On the other hand, if a saccadic movement passes over a cell but there are no fixations inside it, the cell is considered unexplored.

For a fixed length $N$ of the trajectory, we have different mean $\varepsilon$ values within the parameter space.
An efficient trajectory is the one that visits many locations, i.e has a large value for $\varepsilon$.
The efficiency values obtained in the parameter space are depicted in Fig.~\ref{fig:fig_3}.C. As expected, the values corresponding to $\lambda=0$ and any $\delta\theta$ are the same as the ones obtained for $\delta\theta=\pi$ and any $\lambda$ value, since these trajectories emerge from the same distribution. The optimal values are those having a large $\lambda$ value and a small $\delta\theta$.
These are typical of a systematic type of search, in which
very few locations are inspected again. Therefore, the final trajectory explores more locations as compared to the random strategy.

From the study of the relative inter-saccadic angle distribution of the CN cohort, we estimate the corresponding parameters $\lambda$ and $\delta\theta$. 
When placed in the parameter space, we see that the experimental data presents a dispersion in the parameter $\lambda$ ranging from 0.1 to 0.8 (see Fig.~\ref{fig:fig_3}.C). Interestingly, the estimated experimental values for $\delta\theta$ are confined within a range related to small angles, therefore, associated to a persistent movement. We find that the center of mass of these data points, i.e the mean values, is close to 0.4 for $\lambda$, and to $\pi/8$ for $\delta\theta$.

For the estimated values of $\lambda$ and $\delta\theta$, we run the VSM with the corresponding values of $N$ for each trajectory and compare the obtained $\varepsilon$ values. The VSM matches~$78\%$ of the experimental data, as shown in Fig.~\ref{fig:fig_3}.D.
For some points, however the 
model fails to predict the expected $\varepsilon$.
For instance, one participant initiates with a systematic search and later breaks away from that strategy to go for a systematic search truncated in space, i.e without reaching the right end of the image (see Fig.~S2). Our model is not able to reproduce this behavior and, therefore, for the same set of parameters, $\lambda$ and $\delta\theta$, it generates instead a more efficient search.

Interestingly, for the CN cohort we do not find an average behavior that is purely systematic, namely a $\lambda$ value close to unity.
As a consequence, one could ask whether the average searching behavior changes for different searching tasks. 
We analyzed two other experimental cohorts, namely, the \textit{``5-2''} and \textit{``Where's Wally?} (WW)\textit{''} sets, as reported in Ref.~\cite{credidio2012statistical} (See Mat. and Methods). 
The 5-2 experiment consists in finding a target, namely, a single number ``5'', in a regular array of distractors, namely, numbers ``2'', that can be in red or green.
The WW experiment corresponds to scanning images from the famous children's book where the goal is to find \textit{``Wally''}, a character wearing a striped shirt in a crowd of people~\cite{wally}.
A visual search task as the 5-2 experiment, where the target and distractors are placed on a regular lattice, might evoke predominantly systematic search strategies as the regularity of the lattice over which the items are placed serves as a guide to the eye. If indeed this is the case, this cohort should be characterized in the parameter space with large $\lambda$ values and small $\delta\theta$. On the other hand, an experiment as WW, which is composed by a very complex scene with no apparent regularities, might evoke a random strategy, as there is no preferential searching direction. In the parameter space this would be achieved with a small $\lambda$ value and, in principle, any value of $\delta\theta$. Within this hypothesis the CN experiment 
can be viewed as an intermediate configuration
between the aforementioned experiments, as the items are positioned on a lattice but with random displacements.

\begin{figure*}[tbhp]
\centering
\includegraphics[width=1.0\textwidth]{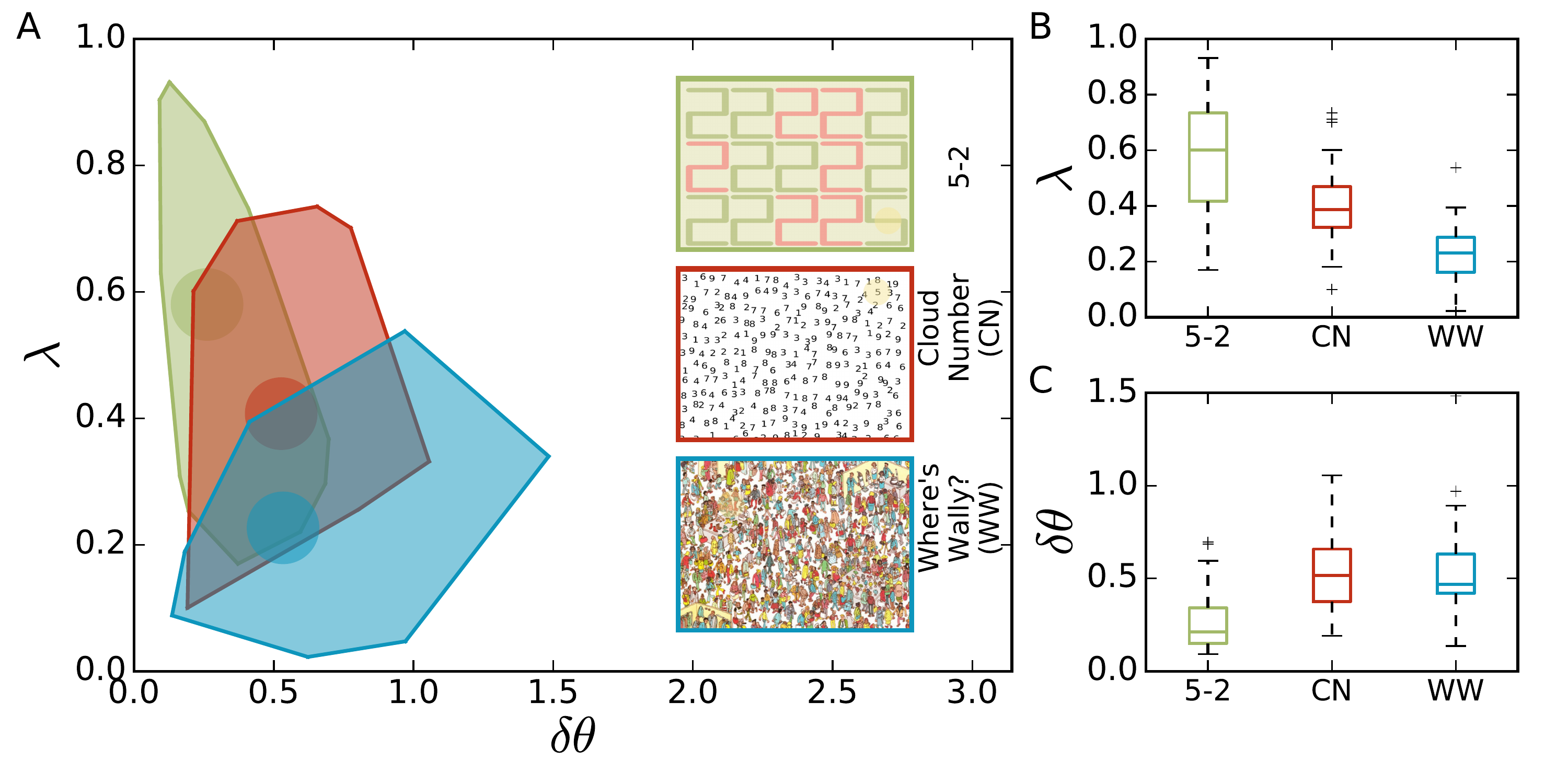}
\caption{\label{fig:fig_4}\textit{Different tasks are placed in distinctive positions in the parameter space.} \textbf{(A)} Three visual search experimental cohorts going from numbers placed in a regular lattice to a random position exhibit distinctive $\lambda$ values. The shaded area shows where the experimental data falls and the circle represents the center of mass of each group. \textbf{(B)} The 5-2 experiment presents a median $\lambda$ value of 0.6 (green), the CN experiment, 0.4 (red), and the WW experiment, 0.2 (blue). \textbf{(C)} For the median value of $\delta\theta$ there is no clear distinction and all experiments are confined to the range [0,$\pi/4$].}
\end{figure*}

We compute the parameters from the 5-2 and WW cohorts (see Fig.S3) and validate the VSM to the experimental data (see Fig.S4). 
The location of all the trajectories composing the experimental cohorts in parameter space are depicted in Fig.~\ref{fig:fig_4}.A. 
The large variability across the $\lambda$ parameter appears within subjects as well as within trials performed by the same subject in a searching episode (see Fig. S5). 
However, even though there is variability present in each of these visual tasks, it is possible to identify a mean strategy employed by each experimental cohort.  
The 5-2 experiment presents a median $\lambda$ value of about 0.6, whereas for the WW experiment it becomes about 0.2, as shown in Fig.~\ref{fig:fig_4}.B. The values for $\delta\theta$ are always confined within [0,$\pi/4$], as shown in Fig.~\ref{fig:fig_4}.C.

By combining the efficiency results from the VSM with the location of the experimental data in parameter space we realize that the 5-2 cohort is the most efficient one. The other trajectories are relatively less efficient. This raises the question whether the efficiency analyzed here, as well as the criteria reviewed in the literature, such as the overall time to find a target, are adequate when it comes to compare different visual search tasks.

Through the study of the inter-saccadic relative angle distribution we were able to quantify the average strategies employed in three visual searching experiments and catalog them according to their $\lambda$ and $\delta\theta$ values. 
The average strategy changes along the different cohorts but always evoking persistent searches (small $\delta\theta$ values), Fig.~\ref{fig:fig_4}.C. One could argue that the discrepancy within the average strategy employed appears as a consequence of the amount of distractors present in the image. However, the 5-2 experiment was performed with different amounts of distractors and 
significant differences between the average $\lambda$ values have not been observed (See Fig.~S6). 
This suggests that the difference between the average strategies appears as a consequence of the scene content and structure.

\section*{Discussion}

Our work focuses on the study of ocular patterns that emerge during a visual search task.
Through
a simple model containing two parameters we show that the knowledge of the saccadic relative orientation distribution gives the relevant information on the visual trajectories found in experiments. The validation of the model with experimental scanpaths through an indirect efficiency measurement confirms this fact.
Moreover, the distribution of the experimental data points over the parameter space shows that indeed there is a large variability between the different visual strategies employed while searching, even for the same subject performing the same experiment. Nevertheless, we are able to discriminate changes in the average strategy across different scenes.
This is shown through the analysis of the model parameters $\lambda$ and $\delta\theta$, which suggests that the visual strategy employed is linked to the underlying structure of the scene, even when the cognitive task remains the same, i.e visual search.

The experimental cohorts presented in this work correspond to participants performing visual searches in different scenarios. Even though these scenarios are easily distinguishable from one another, they have the following common properties: 
a) the scenes are static, no moving items were analyzed in these experiments;
b) the scenes have no salient regions, therefore there are no \textit{a priori} privileged regions to explore in the image; 
c) the task always involves finding a unique target hidden in a set of distractors;
e) the distractors have the same size as the target and,
f) all distractors are equal, in principle, there are no distractors less relevant than others. 
The layout of the items (from a regular lattice to a crowded space) and the features associated to the distractors are the main sources of differences between the experiments.
In the 5-2 experiment, the items can either be a number \textit{``5''} or a number \textit{``2''}. On the other hand, for the WW experiment, the items presented in the scene, although they always involve people, present a variety of shapes and colors that make them richer in visual content, thus, making the scene more complex.

The difficulty of the task in which the participant is engaged while performing visual search has a strong effect on the number of items that can be processed during a fixation~\cite{motter2008changes,geisler1995separation}. Following this idea, it has been suggested that what determines a visual task to be easy or difficult is linked to the discriminability of the target~\cite{young2013eye,hulleman2014search}. Accordingly, we can arrange our set of experiments in the following order of increasing difficulty: 5-2, CN and WW. In this way, our results indicate that a difficult task is more likely to be executed in a random fashion with a relative degree of persistence, whereas an easy task evokes a more systematic search.
An intermediate difficulty task, such as the CN experiment, leads to a mixed type of strategy within the aforementioned extremes.

Our method is simple enough to be used as a tool to further understand or diagnose mental disorders. For instance,
a recent study investigated how the visual patterns, while performing different free viewing tasks, changes for healthy control versus schizophrenic patients~\cite{egana2013small}. The authors presented a set of scenes, with different levels of image complexity, for the participants to freely explore and recorded the position of theirs eyes while doing so. They found that both groups reduced the area of exploration as the images become less complex, while schizophrenic participants tend to maintain the same type of scanpath for all the scenes. Therefore, it would be interesting to measure the parameters $\lambda$ and $\delta\theta$ over these trajectories and study how they change from one group to the other. We believe, based on the reported trajectories in Ref.~\cite{egana2013small}, that the average $\lambda$ should change for the healthy control group, whereas for the schizophrenic group it should remain constant along the different levels of image complexity.  

Our findings and methods can be also applied to other types of search
such as way-finding, where similar patterns appear for people looking for a target in a confined or open space~\cite{pingel2014relationship}. Furthermore, within the framework of foraging and random walks in general, we propose a very simple model that exhibits persistent behavior similar to the effect of ``wind'', namely, a directional bias in a random walk search~\cite{palyulin2014levy}. Additionally, our model comes with the advantage of being able to quantify the type of search by means of two parameters which can be easily determined from experimental data.

In summary, we were able to use the parameter space of the model to correlate a particular scene with a distinct average strategy.
With this new relationship between visual strategies and scene content we can predict 
the average strategy applied in a new experiment by comparing the new scene structure to the ones discussed here. As a perspective work, it would be also interesting to extend our findings to search tasks that involve moving objects, or perhaps to general visual tasks where the sequence of fixations is relevant. Furthermore, given that the scenes are shown on a 2D display with well defined boundaries, it would be worth investigating whether the systematic search strategies remain once the boundaries are removed.

{\small \section*{Materials and Methods}
Both experimental cohorts were conducted in the Universidade 
Federal do Cear\'a with participants who had normal or corrected to normal vision. The experimental data corresponding to the CN cohort is published in Ref.~\cite{amor2016persistence} and the one corresponding the 5-2 and WW cohort, in Ref.~\cite{credidio2012statistical}.
The study has been approved by the Ethics Research Committee of the Universidade 
Federal do Cear\'a (COMEPE) under the protocol number 056/11. All methods used in this 
study were carried out in accordance with the approved guidelines and all experimental 
protocols were approved by COMEPE. Informed consent was obtained from all subjects.

\vspace{0.15cm}
\noindent
{\bf \textit{``Cloud number} (CN)\textit{''} cohort.}
Experimental data was recorded with an EyeLink 1000 system (SR Research Ltd., Mississauga, Canada), 
with an acquisition frequency of 1kHz on a monocular recording over ten subjects (Mean age: 24).
Each subject carried out a sequence of four trials, each one with a maximum duration of five minutes.
In each trial we presented the participant an image with numbers randomly distributed in a $1024\times1280$px image, where the goal was
to find a unique number ``5'' within 1499 distractors.
Between each trial, the subject had the possibility of relaxing and before starting the recording we performed a new calibration.
At the beginning of each trial, the participant was asked to fixate his/her eyes on the center of the screen, 
in case a drift correction needed to be performed.

The classification of the fixation and saccades was made using the EyeLink online filter ~\cite[Section 4.3]{EyeLink}. 
Fixations in the EyeLink system are identified using a saccade-pick algorithm. 
The system analyzes the moment-to-moment velocity and acceleration of the eye using fixed thresholds 
for both eye velocity and acceleration.
If the eye goes above either the velocity or acceleration threshold, the start of a saccade 
is marked.
Analogously, when both the velocity and the acceleration drop back below their thresholds,
the algorithm identifies the saccade end. By default, every movement which does not lie within this definition is considered as being part of a fixation.
The saccade velocity threshold was set to be 30$^\circ/{\rm s}$, the saccade acceleration threshold, 8000$^\circ/{\rm s^2}$, and the saccade motion threshold, 0.15$^\circ$.

\vspace{0.15cm}
\noindent
{\bf{\textit{``5-2''} and \textit{``Where's Wally?} (WW)\textit{''} cohorts.}}
Eye movements were recorded with a Tobii T120 eye-tracking system (Tobii Technology), over 11 subjects (Mean age: 23). The stimuli were presented on a 17” TFT-LCD monitor with resolution $1024\times1280$px and acquisition frequency of 60Hz.
The 5-2 experiment consists on finding a number ``5'' within an array of numbers ``2'' serving as distractors. All numbers (target and distractors) are positioned on a square lattice and are randomly colored red or green, hindering the visual detection of the target through the identification of patterns on the peripheral vision. 
The number of distractors present in each task is related to a degree of difficulty: DF0 (207 distractors), DF1 (857 distractors) and DF2 (1399 distractors). The maximum time given to search the target was 1, 1.5 and 2 minutes respectively. 

The WW experiment consists in finding \textit{``Wally''}, the famous character from the series of books with the same name~\cite{wally}, who is hidden within a very complex background of crowded characters, with a maximum searching time of 2 minutes. 

As explained in Ref.~\cite{credidio2012statistical} the identification of fixations and saccades was carried out with a modified version of the fixation filter developed by Olsson~\cite{ollson2007real}.}

\vspace{0.15cm}
{\small \noindent {\bf{Data availability.}}
The datasets analyzed in the current study are available from the corresponding author upon reasonable request.

\vspace{0.3cm}
{\small \noindent {\bf{Acknowledgments.}}
The authors would like to thank H.F.Credidio for making available the experimental data corresponding to the 5-2 and WW cohorts. This work was supported by the Brazilian agencies CNPq, CAPES, FUNCAP, and National Institute of Science and Technology for Complex Systems in Brazil. We acknowledge financial support from the European Research Council (ERC) Advanced Grant 319968-FlowCCS.}

\vspace{0.3cm}
{\small \noindent {\bf Author contribution.}
T.A.A., M.L., H.J.H. and J.S.A. designed research; T.A.A. performed the experiments; T.A.A. and M.L. analyzed data; T.A.A., M.L., H.J.H. and J.S.A. performed research; T.A.A., M.L., H.J.H. and J.S.A. wrote the paper.}

\vspace{0.3cm}
{\small \noindent {\bf Competing financial interests.}
{\small \noindent The authors declare no conflict of interest.}

\bibliography{references_interface.bib}

\end{document}